# Imaging moiré flat bands and Wigner molecular crystals in twisted bilayer MoTe$_2$

Yufeng Liu[1,2,†], Yu Gu[1,†], Ting Bao[3,4,†], Ning Mao[5], Shudan Jiang[1], Liang Liu[1,2,8], Dandan Guan[1,2,8], Yaoyi Li[1,2,8], Hao Zheng[1,2,8], Canhua Liu[1,2,8], Kenji Watanabe[6], Takashi Taniguchi[7], Wenhui Duan[4], Jinfeng Jia[1,2,8], Xiaoxue Liu[1,2,8], Can Li[1,2,*], Yang Zhang[3,9,*], Tingxin Li[1,2,8,*], and Shiyong Wang[1,2,8,*]

[1]Tsung-Dao Lee Institute, Shanghai Jiao Tong University, Shanghai, 201210, China

[2]Key Laboratory of Artificial Structures and Quantum Control (Ministry of Education), School of Physics and Astronomy, Shanghai Jiao Tong University, Shanghai 200240, China

[3]Department of Physics and Astronomy, University of Tennessee, Knoxville, TN 37996, USA

[4]Department of Physics, Tsinghua University, Beijing 10084, China

[5]Max Planck Institute for Chemical Physics of Solids, 01187, Dresden, Germany

[6]Research Center for Electronic and Optical Materials, National Institute for Materials Science, 1-1 Namiki, Tsukuba 305-0044, Japan

[7]Research Center for Materials Nanoarchitectonics, National Institute for Materials Science, 1-1 Namiki, Tsukuba 305-0044, Japan

[8]Hefei National Laboratory, Hefei 230088, China

[9]Min H. Kao Department of Electrical Engineering and Computer Science, University of Tennessee, Knoxville, Tennessee 37996, USA

[†]These authors contribute equally to this work.
[*]Emails: lic_18@sjtu.edu.cn, yangzhang@utk.edu, txli89@sjtu.edu.cn, shiyong.wang@sjtu.edu.cn

**Abstract**

**Two-dimensional semiconducting moiré materials have emerged as a highly tunable platform for exploring novel quantum phenomena such as interaction-driven electronic crystals, correlated topological phases, and unconventional superconductivity[1–18]. Recently, twisted bilayer MoTe$_2$ (tMoTe$_2$) has attracted significant attentions due to the observation of the long-sought fractional quantum anomalous Hall effect[19–34]. However, a comprehensive microscopic understanding of the tMoTe$_2$ moiré superlattice remains elusive. Here, we report scanning tunneling microscopy/spectroscopy (STM/STS) studies in dual-gated tMoTe$_2$**

abstract
**moiré devices with twist angles ranging from 2.3° to 3.8°. The device consists of two independent back-gates: one enables an ohmic contact for tMoTe$_2$, while the other fine-tunes the Fermi level of tMoTe$_2$. This dual-gate control enables direct measurement of the electronic structure in tMoTe$_2$ under varied displacement fields and moiré filling factors, by fine tuning the gate voltage and the tip bias. Our STS spectra and spatial imaging reveal that the low-energy moiré flat bands are predominantly localized in the XM and MX regions of the moiré superlattice. At zero electric field, these bands form a honeycomb lattice with non-trivial topology, whereas an applied electric field drives a transition into two distinct triangular lattices with trivial topology. The spatial distributions align with large-scale first-principle calculations, demonstrating that the topological flat bands arise from the K-valley hybridization between the top and bottom MoTe$_2$ layers. Furthermore, we show that the effective moiré potential depth can be electrostatically controlled via gate and tip biases. At sufficient potential depths, we observe the emergence of Wigner molecular crystals, transitioning MX triangular lattice into a Kagome lattice at MX moiré filling factor $v_{MX}$ = 3. These results elucidate the microscopic origin of topological flat bands in tMoTe$_2$ and demonstrate electric-field control of topology and correlated electronic orders, paving the way to engineer exotic quantum phases in moiré simulators.**

**Introduction**

Moiré flat bands in semiconducting transition metal dichalcogenide (TMDc) systems can be well described by a two-dimensional electron gas under a modulated moiré potential[35,36]. In TMDc moiré heterobilayers and twisted homobilayers with layer asymmetry, these isolated low-energy flat bands effectively emulate Hubbard physics on a triangular lattice[35,36]. Conversely, in band-aligned moiré bilayer, nontrivial band topology can emerge from the hybridization of two sets of moiré flat bands from different layers, forming a honeycomb lattice[5,6]. In real space, the low-energy physics can be understood in terms of carrier hopping on a layer-pseudospin skyrmion lattice, where the layer polarization of the wavefunctions is spatially modulated[5-7]. To date, topological flat bands have been experimentally realized in systems such as WSe$_2$/MoTe$_2$ heterobilayers[9,10], tMoTe$_2$[19–27], and twisted bilayer WSe$_2$[11,12]. These systems exhibit a wealth of topological quantum phases, such as the integer and fractional quantum anomalous Hall effects.

Recent theoretical work further predicts that intra-moiré interactions can be sufficiently strong to stabilize a Wigner molecular state in TMDc[37–39], leading to a transition from a triangular lattice to an emergent Kagome lattice at a moiré filling factor of $v$ = 3. While TMDc moiré flat bands have been directly visualized in previous STM/STS studies[40–51], direct atomic-scale characterization under in situ tunable displacement fields and filling factors has remained elusive. This gap hinders a unified understanding of the interplay between topology, interactions, and external tuning parameters—key to unlocking the full potential of moiré quantum simulators.

Here, we performed STM/STS investigations on tMoTe$_2$ with twist angles ranging from 2.3° to 3.8°. Although tMoTe$_2$ is the only TMDc material known to exhibit a fractional quantum anomalous Hall (FQAH) state, STM/STS studies of this system have been scarce due to its instability under ambient conditions. To address this challenge, we employed a monolayer of hexagonal boron nitride (h-BN) as a protective layer, effectively preserving the intrinsic properties of MoTe$_2$ while enabling electron tunneling (Methods). Furthermore, we developed a device structure with two independent back gates, allowing us to directly measure the electronic structure of h-BN-encapsulated tMoTe$_2$ under tunable displacement electric fields by fine tuning the gate voltage and the tip bias. Notably, the tunability of the moiré bands through displacement fields provides a critical pathway for unraveling the influence of electron correlations and band topology in this material. Our STS spectra and spatial imaging reveal that the low-energy moiré flat bands, predominantly localized at XM and MX regions, form a honeycomb lattice under zero electric field. These spatial distributions are consistent with first-principles calculations, which attribute the flat bands to K-valley band hybridization between the top and bottom MoTe$_2$ layers. By controlling the moiré potential depth via a perpendicular electric field, we observed Wigner molecular crystals transitioning from a triangular lattice to a Kagome lattice at $v_{MX}$=3. These results provide real-space evidence of lattice reconstructions and topological moiré flat bands in tMoTe$_2$ and their modulations under displacement electric fields and filling factors.

**Two setups for the microscopic characterization of t-MoTe$_2$ moiré superlattice**

When two layers of MoTe$_2$ are twisted by a small angle, the long period moiré superlattice is formed with three high-symmetry stackings, denoting as MM, MX, and XM, as depicted in Fig. 1a and 1b. Although the effect of period moiré potential can be understood by considering a rigid lattice structure with continuous variation in stacking configurations, the atomically inter- and intra-lattice reconstructions can profoundly influence the moiré band structure. We performed large-scale DFT calculations to address the lattice relaxation effect of tMoTe$_2$ with a twist angle of $\theta$ = 2.65° and 3.89°. The calculation results reveal an out-of-plane corrugation induced by inter-lattice interactions (Fig. 1c), with the MM region elevated by 0.42 Å compared to the XM region, and the XM regions is slightly lower by 0.02Å than the MX regions. Additionally, the intralayer strain exhibits a helical chirality, leading to the generation of a substantial pseudomagnetic field. The calculated band structure including the lattice reconstruction effects is shown in Extended Data Fig. 1.

Figure 1d and Figure 1h illustrate the schematic configurations of the tMoTe$_2$ device used for STM measurements. Because tMoTe$_2$ is an insulator—making direct STM measurements unfeasible—we designed two distinct device configurations. In the first setup (Fig. 1d–e), the tMoTe$_2$ flake is placed directly on a few-nanometer-thick exfoliated graphite substrate that serves both as the electrode and as the conductive platform for tunneling measurements. This device architecture has been widely employed in previous STM/STS studies of semiconducting TMDs and their moiré systems. However, in this configuration, the Fermi level of tMoTe$_2$ remains locked within the semiconducting gap, pinned by the graphene substrate, which restricts the exploration of its electronic structure under varying filling factors. In the second setup (Fig. 1h-i), a dual back-gating strategy is implemented: the global Si/SiO$_2$ gate induces heavy hole doping to establish an ohmic contact between tMoTe$_2$ and the platinum electrodes, while the local graphite gate independently controls the Fermi level of tMoTe$_2$ under STM tip. This device structure enables reliable STM/STS measurements of tMoTe$_2$ across a range of displacement electric fields and doping densities.

STM measurements were conducted on five tMoTe$_2$ devices with slightly different twist angles, ranging from approximately 2.3° to 3.8° (Extended Data Fig. 6-7). STM imaging reveals micrometer-size clean area with typical uniaxial strain up to 2% (Extended Data Fig. 2). As shown in Fig. 1f and Fig. 1j, the STS differential conductance (d$I$/d$V$) spectra reveal the semiconducting gap of tMoTe$_2$. Specifically, the valence band (VB) top is located at -0.8 V away from the Fermi level as probed by the Setup 1 (Fig. 1f), while another is aligned close the Fermi level by the setup 2 (Fig. 1j). The measured band gap is approximately 1.6 eV. The STM topographic imaging reveals structural features that closely resemble the lattice reconstructions predicted by DFT calculations. Notably, the moiré superlattice exhibits XM/MX regions with reduced height, which are separated by domain walls and MM regions (Fig. 1g and Fig. 1k).

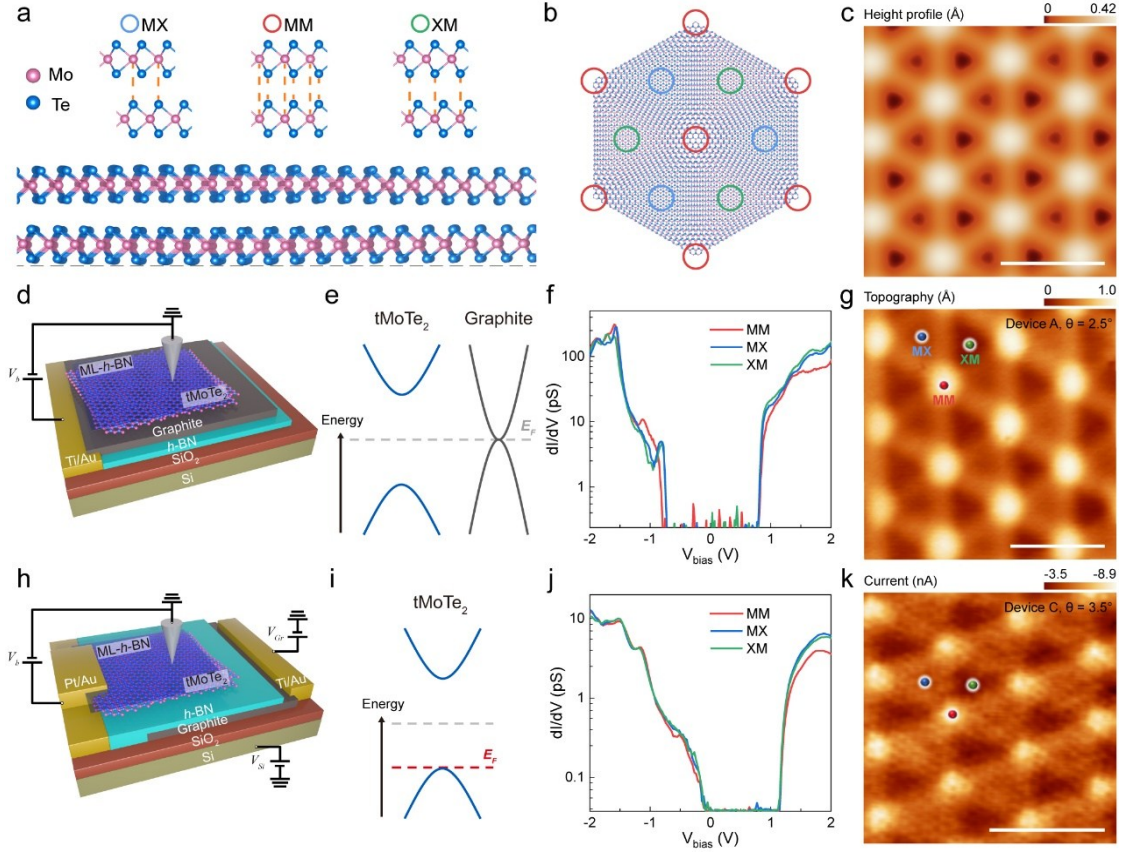

**Fig. 1. Device structure and lattice relaxation of tMoTe$_2$. a,b,** Side and top view of the schematic moiré superlattice of tMoTe$_2$. High-symmetry stackings (MX, XM, MM) are highlighted by circles. **c,** DFT calculated height profile for **2.65°** tMoTe$_2$. **d,** Schematic of the **Setup 1** tMoTe$_2$ device for STM measurements. A monolayer of h-BN is used to protect the air-sensitive MoTe$_2$ flake while facilitating electron tunneling via the supporting thin graphite film. **e,** Schematic of band alignments between tMoTe$_2$ and graphite. **f,** STS d$I$/d$V$ spectra taken at the sites marked in **g**. (Setup 1 device A; Bias Modulation: 20 mV). **g,** Experimental STM topography of ~**2.5°** tMoTe$_2$ (Setup 1 device A: $V_{bias}$ = -1.5 V, I = 1 nA). **h,** Schematic of the **Setup 2** tMoTe$_2$ device for STM measurements. The silicon gate is used to establish ohmic contact between MoTe$_2$ and the electrodes, while the graphite gate is used to tune the filling of tMoTe$_2$. **i,** Schematic of the filling of tMoTe$_2$ bands during measurements. **j,** STS d$I$/d$V$ spectra taken at the sites marked in **k**. (Setup 2 device C; Bias Modulation: 40 mV). **k,** Experimental current topography of ~**3.5°** tMoTe$_2$ (Setup 2 device C: $V_{bias}$ = -0.76 V).

**Probing the low-lying Moiré Flat Bands at K and Γ Valleys**

Band folding induced by the moiré superlattice creates flat electronic bands in the mini-Brillouin zones. Previous transport and optical measurements suggest that the moiré flat bands at the ±K valleys are closer to the Fermi level than those at the Γ valley[19-22]. To probe the electronic structure and disentangle the contributions from the K and Γ valleys, we performed DFT calculations and STS measurements. As shown in Fig. 2b, the moiré flat bands originating from the K and Γ valleys in tMoTe$_2$ are illustrated, with the K-valley bands situated nearer to the Fermi level. Figures 2c and 2d present the density of states (DOS) spectra for the K and Γ valleys, performed for twist angles of 2.65° and 3.89°, respectively. The energy separation between the K- and Γ-valley bands decreases as the twist angle is reduced, underscoring the sensitivity of the electronic structure to the twist configuration.

Our calculations reveal that the states at the Γ-valley arise primarily from the *dz²* orbitals of Mo and Te, which exhibit significant out-of-plane orbital character and decay slowly outside the MoTe$_2$ layer (Extended Data Fig. 3). In contrast, the K states originate from the in-plane $d_{xy}$ orbitals of Mo and the $p_x/p_y$ orbitals of Te. These wavefunctions, characterized by strong in-plane orbital character, decay rapidly outside the MoTe$_2$ layer. This rapid decay of K-valley states poses challenges for STS measurements, requiring a shorter tip-sample distance for enhanced resolution.

In Setup 1, the VB top is located far from the Fermi level (approximately -0.8 eV), necessitating a relatively large tip–sample distance of about 1 nm to maintain a stable tunneling condition. However, this larger separation limits the STM's energy resolution, making it difficult to resolve the fine features of the moiré flat bands associated with the K valley. Consequently, with Setup 1 we observe only a modest increase in the local density of states (LDOS) at the MX/XM regions compared to the MM regions within a bias range of -0.7 V to -0.9 V (Fig. 2e). Spatial STS mapping further reveals that the LDOS is predominantly concentrated at the MX regions, with the area of high intensity expanding as the bias increases (Fig. 2f). These spatial distributions are consistent with DFT-calculated DOS maps using the states of K-valley moiré bands from the top MoTe$_2$ layer (Fig. 2g). Given that STM is a surface-sensitive technique, the tunneling signal is dominated by the electronic states of the top layer, especially at the larger tip–sample distances employed in this configuration. As a result, only a single triangular lattice corresponding to the top layer is resolved in Setup 1.

In Setup 2, the VB top is tuned near the Fermi level, enabling STM measurements at a significantly reduced tip–sample distance and thereby enhancing LDOS intensity. This improvement facilitates the clearer detection of the K-valley electronic states from both the top and bottom layers. As shown in Figure 2h, well-defined LDOS peaks are observed at both MX/XM and MM sites. Notably, the peak at the MX/XM sites appears at a lower energy than that at the MM sites, in agreement with the DFT calculations presented in Figure 2c. Furthermore, our d$I$/d$V$ spectroscopy performed at varying tip–sample distances reveals that the low-lying states at the XM/MX sites decay much more rapidly with increasing tip-sample separation compared to the states at the MM sites (Extended Data Fig. 3). These results confirm that the moiré flat bands originating from the K valley are situated above those derived from the Γ valley, thereby confirmng the low-energy physics in twisted MoTe$_2$ is dominated by the K-valley moiré bands.

Spatially resolved STS measurements further reveal the intricate distribution of K-valley states under an applied displacement electric field of $V_{Gr}$ = -6 V (Fig. 2i). At a tip bias of -57 meV, a distinct triangular lattice pattern emerges at the MX sites of the top layer. As the tip bias increases to -107 meV, an additional triangular lattice becomes evident at the XM sites of the bottom layer. At -140 meV, the dot-like features evolve into well-defined ring-like structures. Under the applied electric field, the degeneracy between the MX and XM moiré bands is lifted, resulting in the formation of two distinct triangular lattices with trivial topology. Furthermore, as shown in Fig. 2j, we simulated the DOS distribution for the top layer at a twist angle of 3.89°. The simulated maps successfully reproduce the evolution from dot-like to ring-like features in the top triangular lattice.

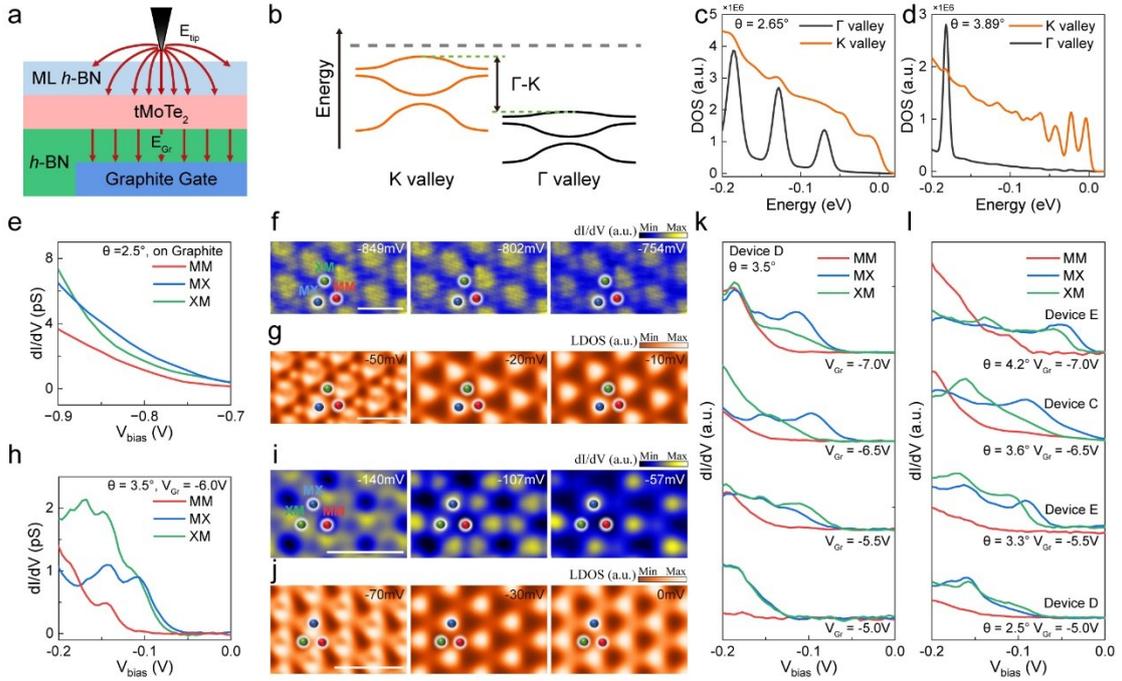

**Fig. 2. The moiré flat bands from K and Γ valleys of tMoTe$_2$. a,** Schematic of the **Setup 2** device. Both the STM tip and the graphite gate introduce displacement fields. **b,** Schematic of band alignments of moiré flat bands from the K and Γ valleys. The moiré flat bands from the K valley lie closer to the Fermi level than those from the Γ valley. **c, d,** DFT-calculated DOS spectra of moiré flat bands for **2.65°** and **3.89°** tMoTe$_2$. **e,** STS d$I$/d$V$ spectra taken at high-symmetry points for a tMoTe$_2$ region with θ ≈ **2.5°** (**Setup 1, Device A**), showing weak features from K-valley moiré flat bands. **f,** Constant-height STS mappings taken at different $V_{bias}$ (Scale bars: 10 nm; Bias Modulation: 20 mV). **g,** DFT-calculated DOS maps of the top layer of **2.65°** tMoTe$_2$ at different energies. **h,** STS d$I$/d$V$ spectra taken at a high-symmetry point for a tMoTe$_2$ region with θ ≈ **3.5°** (**Setup 2, Device D**), showing clear peak features from K-valley moiré flat bands. The moiré flat bands from the K valley are barely visible in d$I$/d$V$ spectroscopy using the **Setup 1** device but are significantly enhanced in **Setup 2** due to reduced tip-sample separation. **i,** Constant-height STS mappings taken

at different $V_{bias}$ (Scale bars: 10 nm; Bias Modulation: 6 mV). **j,** DFT-calculated DOS maps of the top layer of **3.89°** tMoTe$_2$ at different energies. **k,** STS d$I$/d$V$ spectra taken at a high-symmetry point for a tMoTe$_2$ region with $\theta \approx 3.5°$ under different graphite gate voltages (**Setup 2, Device D**), indicating that the **Γ-K** separation depends sensitively on displacement fields. **l,** STS d$I$/d$V$ spectra taken at a high-symmetry point for different tMoTe$_2$ regions with varying twist angles and graphite gate voltages.

Figures 2k and 2l present the d$I$/d$V$ spectra for devices 3 and 4, measured under varying twisted angles and graphite gate voltages. Across all examined gate voltages, the moiré flat bands associated with the K valley consistently remain closer to the Fermi level than those originating from the Γ valley. Rather than exhibiting a gradual spectral shift as predicted by a single-particle model, we observe pronounced variations in the energy separation between the K- and Γ-valley bands, in the peak intensities, and in the positions of the states at the MX/XM regions as the applied electric field is varied. These observations indicate that the moiré flat bands are highly tunable by displacement electric field. We observe that STS energy resolution is also influenced by back-gate displacement fields. At low graphite gate voltages, the MX/MX bands are degenerate, and clear isolated peaks corresponding to different moiré flat bands become visible (Extended data Fig. 4).

**Modulating K-valley Moiré Flat Bands by Displacement Electric Field**

The application of a displacement electric field introduces potential energy differences between the top and bottom MoTe$_2$ layers, significantly altering the interlayer coupling and the shape and depth of the moiré potential. Figure 3a illustrates the Brillouin zones of t-MoTe$_2$, with the regions outlined by red and blue lines representing the top and bottom layers, respectively. Due to the small twist angle, the Brillouin zones of the two layers are slightly rotated relative to each other in reciprocal space. At zero displacement field, the flat bands at the K valley of the top layer (K$_T$) are degenerate with those of the bottom layer (K$_B$) (as marked by dashed grey curves in Fig. 3b). The hybridization of two sets of moiré flat band from different layer effectively mimics the Kane-Mele model on a honeycomb lattice, enabling the observation of integer and fractional quantum anomalous Hall effects. However, applying a displacement electric field lifts the degeneracy between the K$_T$ and K$_B$ bands, as shown in Figs. 3b and 3c. At positive displacement fields, the K$_T$ bands move closer to the Fermi level, while the K$_B$ bands are pushed farther away (Fig. 3b). Conversely, at negative displacement fields, the K$_T$ bands are pushed farther from the Fermi level, and the K$_B$ bands shift closer (Fig. 3c).

To investigate these effects, we performed dense d$I$/d$V$ spectra along the high symmetry direction while varying the graphite gate voltage to tune the out-of-plane displacement field (Figs. 3d–f, Extended Data Fig. 4). In setup 2, the displacement electric field $D$ is affected by two factors: one from the STM tip (tip bias and tip work function) and the other from the graphite back gate (as illustrated in Fig. 2a). At -5.5 V graphite gate, the tip-induced *D* and the gate-induced *D* compensate and result in a nearly zero net displacement field. As shown in Fig. 3d, the $K_T$ and $K_B$ bands are nearly degenerate, forming a honeycomb lattice that persists across different bias voltages (Fig. 3g). Increasing the gate voltage to -6.5 V introduces a positive net displacement field, shifting $K_T$ bands closer to the Fermi level while pushing $K_B$ bands further away. This breaks the honeycomb symmetry, with the $K_T$ bands becoming dominant and forming a triangular lattice at lower biases (Fig. 3h). At an even larger displacement field (e.g., -7.0 V graphite gate voltage), the $K_T$ and $K_B$ bands remain non-degenerate and exhibit triangular lattice patterns in the -50 meV imaging, as shown in Fig. 3i. Notably, at such high displacement electric fields, the lowest-energy states are not localized at the center of the MX/XM regions, as typically expected, but instead appear near the potential edges, closer to the domain walls (marked by dashed circles in Fig. 3f). To provide clearer insight into these distributions, we marked the MX/XM centers with colored dots, highlighting that the maximum intensities of the states are shifted away from these centers (Fig. 3i). Such distributions suggest the shape of moiré lattice may alter under strong displacement fields.

We calculated the moiré band structure and DOS spectra using a continuum model that incorporates the effects of an applied displacement electric field (Extended Data Fig. 5). This relatively simple model successfully captures the key experimental findings. At zero field, the DOS spectra corresponding to the XM and MX states are identical and lie at lower energies than those of the MM sites. As the applied displacement field increases, the degeneracy between the XM and MX bands is progressively lifted, resulting in an energy splitting between the corresponding DOS peaks. Moreover, the magnitude of this splitting grows with increasing electric field, in agreement with our experimental observations.

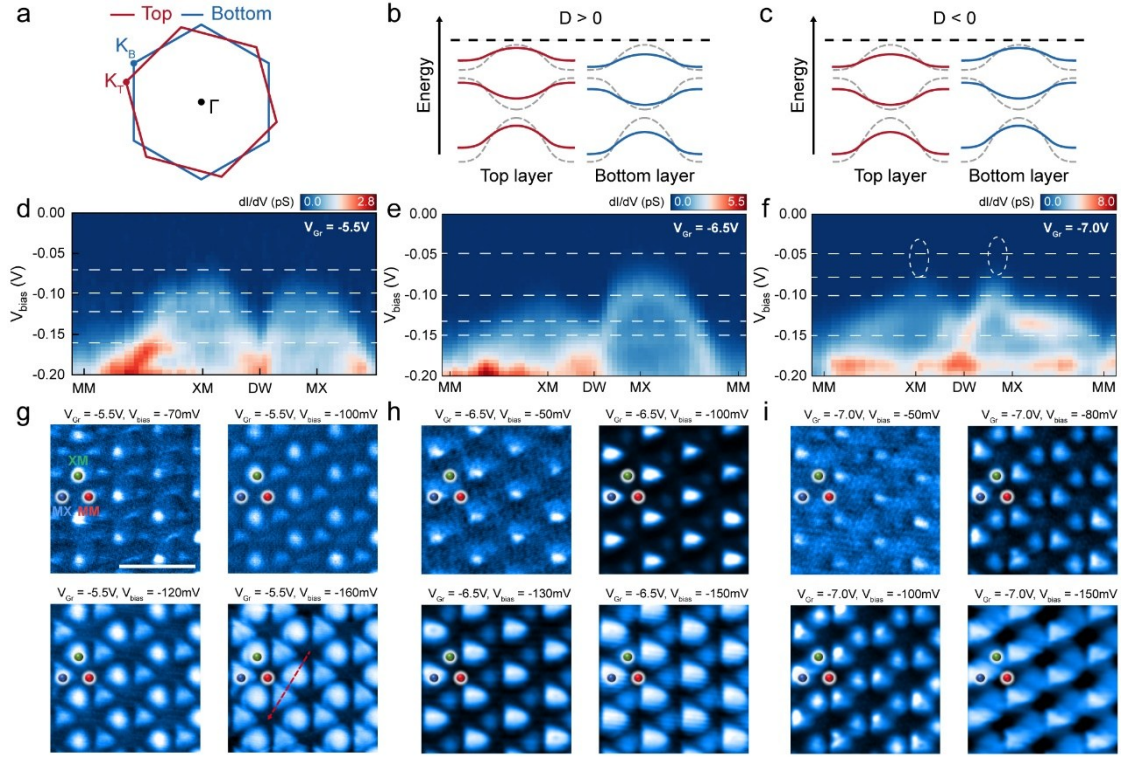

**Fig. 3. Modulating moiré flat bands from K valleys by displacement electric fields. a,** Schematic of the Brillouin zone of tMoTe$_2$, with regions outlined by red and blue lines representing the top and bottom layers, respectively. **b, c,** Schematic of band alignments of moiré flat bands from the K valleys under positive and negative displacement electric fields, respectively. At zero displacement field, the flat bands at the K valley of the top layer ($K_T$) are degenerate with those of the bottom layer ($K_B$) (dashed gray curves). A displacement electric field lifts the degeneracy between the $K_T$ and $K_B$ bands, shifting their relative energies and modifying the band structure. **d-f,** STS d$I$/d$V$ spectra measured along the dashed line in **g** at different graphite gate voltages (**Setup 2, Device D**), demonstrating the tunability of the moiré flat bands via displacement fields. **g-i,** Constant-height current images taken at different $V_{bias}$, showing spatial variations in the local density of states (Scale bars: 10 nm). These images highlight the influence of displacement fields on the electronic structure within the XM/MX moiré superlattice.

**Emergent Wigner Molecular Crystals in Twisted MoTe$_2$**

The electron density distributions inside moiré potential can be imaged via tunneling through the valence band edges[51,52]. Using this approach, we examined the electron density distributions at the MX and XM moiré sites under an applied displacement field. Under conditions of strong electron-electron correlations, we observed behavior characteristic of Wigner molecules at the MX moiré sites. The concept of Wigner molecules arises from the interplay of Coulomb repulsion and quantum confinement. Strong electron-electron interactions lead to localized charge distributions, with electrons occupying distinct positions within the confined region.

The Wigner parameter $R_W=U/\Delta$ quantifies the competition between Coulomb repulsion energy ($U$) and single-particle energy-level spacing ($\Delta$). In moiré lattices, the displacement field and material properties determine the depth of the moiré potential, setting the energy scale of $\Delta$, while the electron density and dielectric environment control $U$. At small $R_W$, the kinetic energy dominates, and electrons occupy quantum orbitals in a non-interacting sequence, with charge density peaking at the center of the confinement potential (Fig. 4c). As $R_W$ increases, Coulomb interactions become dominant, causing electrons to spatially separate and localize at distinct positions to minimize mutual repulsion (Figs. 4a-c). By tuning the displacement field, $R_W$ can be controlled in tMoTe$_2$, enabling manipulation of the triangular XM/MX moiré lattices.

Valence-band-edge tunneling maps (Figs. 4d-f) illustrate the evolution of electron configurations in moiré artificial molecules as the electron occupancy per MX site ($v_{MX}$) increases. For $v_{MX}=1$, a single charge density node appears at each moiré site center (Fig. 4d), corresponding to a strongly localized electron within the deep moiré potential. At $v_{MX}=2$, two distinct charge nodes are observed at MX moiré sites due to Coulomb repulsion exceeding the single-particle energy gap, resulting in spatial separation of the two electrons. When $v_{MX}=3$, the charge density evolves into a trimer configuration at MX moiré sites, with three distinct peaks arranged in a triangular pattern. Notably, the charge density exhibits a local minimum at the center of the site, despite the lowest potential being located there (Fig. 4e). This configuration is a direct visualization of Wigner molecule formation, where Coulomb repulsion dominates over the single-particle energy gap.

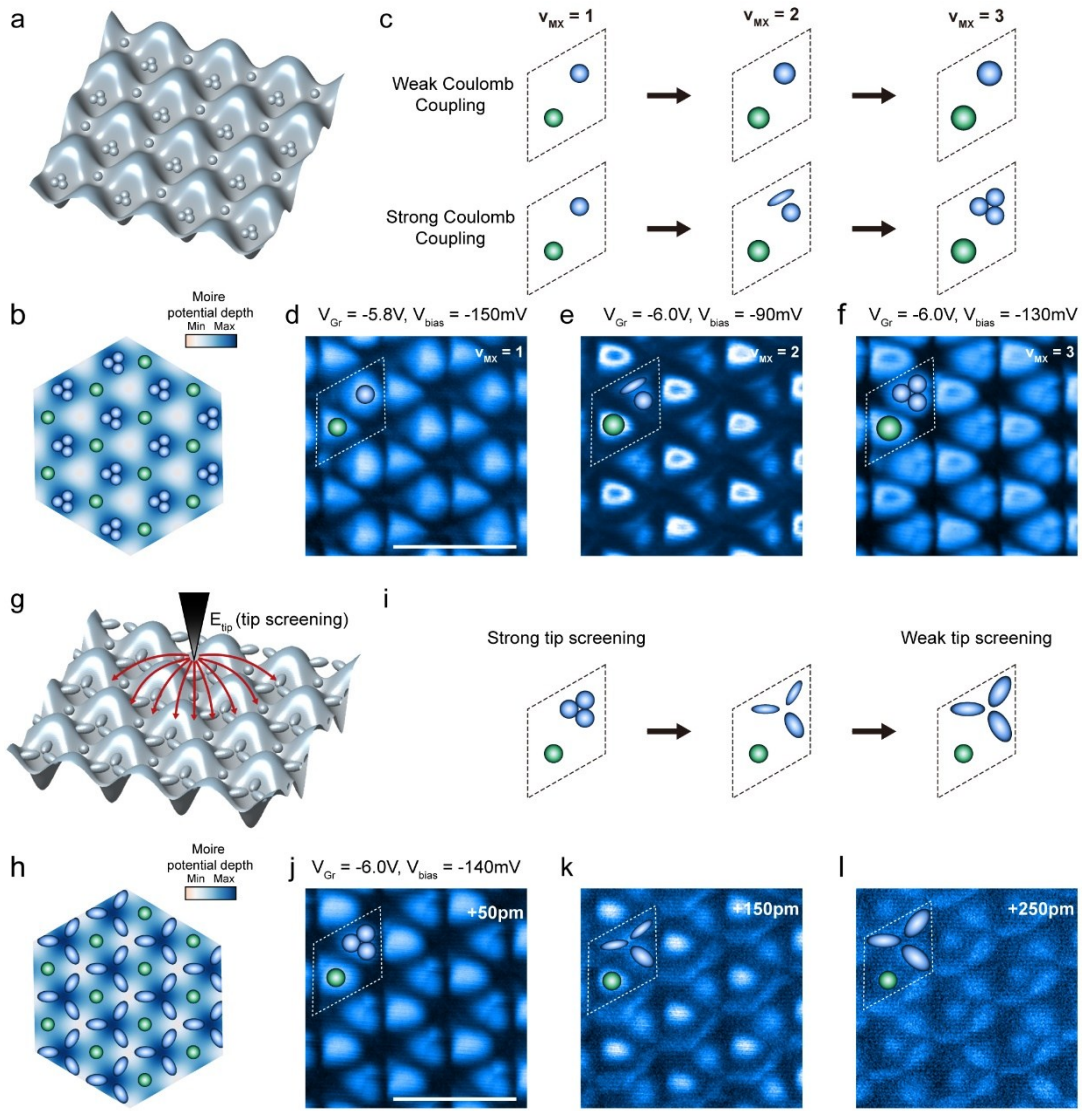

**Fig. 4. Visualizing and tuning Wigner molecular crystals. a, b,** Schematic of Wigner molecular crystals at a filling of $v_{MX} = 3$ per moiré site. Under displacement electric fields, the **MX** moiré potential becomes deeper than the **XM** moiré, leading to stronger Coulomb interactions. At $v_{MX} = 2$ and $v_{MX} = 3$ fillings, electrons spatially separate and localize at distinct positions to minimize mutual repulsion, forming a Wigner molecular crystal. **c,** Schematic illustration of electron positions under weak and strong Coulomb interactions. As Coulomb interactions weaken, the Wigner molecule transitions toward a more delocalized state, resembling conventional electronic structures with overlapping wavefunctions. **d-f,** Valence-band-edge tunneling maps showing electron densities at different biases and graphite gate voltages. These maps reveal the spatial arrangement of localized electrons and how their distribution evolves under external tuning. (**Setup 2, Device D**). **g,** Schematic of the STM tip screening effect, where the local electric field from the tip influences electron localization. **h,** Schematic of electron density distribution under weak tip-screening conditions, showing enhanced delocalization of electrons inside **MX** moiré sites. **i,** Schematic of the reduced tip-screening effect, which enhances electron delocalization. **j-l,** Valence-band-edge tunneling maps showing electron densities at different tip-sample separations. The STM tip was gradually retracted from the initial setpoint (**Setup 2, Device D**, $V_{bias}$ = -140 mV, **I** = 18 pA, at **XM** sites), demonstrating the effect of tip-induced perturbation on electron localization. Scale bars: 10 nm.

The dielectric environment plays a critical role in governing Coulomb repulsion between electrons in moiré superlattices. In tMoTe$_2$, the effective dielectric constant can be tuned by adjusting the tip-sample distance in the STM junction, which modulates tip-induced screening and, consequently, the strength of Coulomb interactions. This tunability facilitates the exploration of transitions between lattice geometries under different Coulomb interactions. As illustrated in Figs. 4g-i, increasing the tip-sample distance weakens the screening effect, thereby enhancing Coulomb repulsion. This increased repulsion drives greater spatial separation of the trimer electrons, ultimately leading to a nearly perfect Kagome lattice at MX moiré sites, while XM moiré sites retain a triangular lattice configuration (Fig. 4h). These tip-induced behaviors have been confirmed by valence-band-edge tunneling experiments: at short tip-sample distances (Fig. 4j), the trimer remains tightly packed at the moiré center; at medium distances (Fig. 4k), the trimer begins to separate into an irregular three-arm configuration; and at larger distances (Fig. 4i), the trimer achieves further separation, forming a highly ordered Kagome lattice at MX moiré sites.

Wigner molecular crystals are periodic arrangements of Wigner molecules, where localized multi-electron states self-organize into ordered patterns due to the interplay between Coulomb repulsion and the periodic moiré potential. For $v_{MX}$ =3, electrons form molecular states within each moiré unit cell, with their periodic positioning at MX sites giving rise to a Kagome lattice. These Kagome crystals should exhibit distinct band structures, making them promising platforms for studying correlated quantum phases and exotic electron ordering[37–39].

**Conclusions and Outlook**

Our study provides a comprehensive atomic-scale investigation of the electronic properties of twisted MoTe$_2$, revealing the intricate interplay between electron correlations, moiré potential, and displacement electric field. By utilizing a protective h-BN monolayer and a dual-gating technique, we achieved precise control over the electronic structure, enabling the direct observation of low-energy moiré flat bands and the field-induced emergence of Wigner molecular crystals. These results establish a microscopic framework for understanding moiré flat band topology and interaction-driven phases in tMoTe$_2$, while highlighting displacement fields as a critical knob for engineering emergent quantum phases. Looking ahead, this work provides a roadmap for probing higher-order moiré fillings and their potential link to non-Abelian states and other novel interaction-driven quantum matter.

**Additional Notes**: During the preparation of this manuscript, we become aware of a recent work that also reports the STM/STS study on tMoTe$_2$[53].

## Methods

**Device fabrications**

The tMoTe$_2$ stacks for STM/STS measurements were fabricated using the standard try transfer method[54] (Extended Data Fig. 1) in a nitrogen-filled glovebox to prevent the degradation of the MoTe$_2$. First, we prepared electrodes (Ti/Au 10 nm/60 nm) on Si/SiO$_2$ substrates using e-beam lithography and e-beam evaporation. Then, monolayer hBN, monolayer MoTe$_2$, graphite (~ 3-5 nm), and hBN (~10-20 nm) were mechanically exfoliated onto Si/SiO$_2$ substrates and identified via optical contrast. A monolayer hBN was initially picked by a polycarbonate (PC) film stamp. This monolayer hBN was then used to picked up half of the monolayer MoTe$_2$ (which was cut into two pieces with an AFM tip before stacking). The remaining MoTe$_2$ was picked up after rotating it by a small angle controlled by a mechanic rotator. After picking up the bottom graphite and bottom hBN layers, the entire stack was released onto the prepatterned Ti/Au electrodes. The monolayer hBN served as the protection layer for tMoTe$_2$. The bottom graphite served as the tunneling electrodes for the STM tip, while the bottom hBN layer screened inhomogeneities generated by substrate surface roughness. The finished stack was dipped in chloroform, acetone and isopropanol for a few minutes, respectively, to remove the majority of the PC film. Finally, any residual contaminants were cleaned by repeated scanning with an atomic force microscope operated in contact mode (Park NX7, setting contact force ~ 300 nN), in order to obtain an atomically-clean surface for STM/STS measurements. Before measurement, these samples were annealed in ultrahigh vacuum at 150 °C overnight.

To protect the air-sensitive tMoTe$_2$ flakes from degradation while maintaining compatibility with STM imaging, we utilize a monolayer of h-BN as a capping layer. The encapsulation with h-BN not only preserves the structural and electronic integrity of the tMoTe$_2$ but also provides an atomically flat and clean surface essential for high-resolution imaging. The detailed fabrication process of these devices is illustrated in Extended Data Fig. 6-7. To ensure an atomically clean surface for STM imaging, we adopt a contact-mode atomic force microscopy scanning procedure. This step effectively removes any residual contaminants from the h-BN surface. Post-cleaning, micrometer-sized clean regions suitable for STM studies are observed, as shown in Extended Data Fig. 2. These setups and preparation techniques enable precise characterization of the electronic structure and moiré patterns in tMoTe$_2$ with high spatial resolution.

**Angle inhomogeneity and strain**

The local twist angle and strain of tMoTe$_2$ are determined using a uniaxial hetero-strain model[55], which assumes that one layer is uniaxially strained by a percentage ε at an angle $\theta_s$ to one of the MoTe$_2$ crystal lattices, while the other layer remains unstrained but with a twist angle $\theta_T$ relative to the first. This model provides three degrees of freedom for the tensile moiré geometry: the moiré twist angle $\theta_T$, the heterostrain magnitude ε, and the angle of uniaxial strain application $\theta_s$. These three variables can be numerically solved by fitting with the experimentally measured three side moiré wavelengths. In this work, the MoTe$_2$ Poisson ratio is estimated to be δ = 0.24[56]. Additionally, a statistical analysis was conducted on two samples, and the respective twist angle ranges and strain levels are illustrated in Extended Data Fig. 2.

**STM/STS measurements**

STS/STM were carried out using a commercial Unisoku Joule-Thomson STM, a commercial Unisoku 1200JT STM system and a commercial CASAcme cryogen free LT SPM under low temperature (4.3K) and ultra-high vacuum conditions ($3 \times 10^{-10}$ mbar). The tungsten tips were calibrated against the surface state of a Cu (111) single crystal or an Ag (111) single crystal. A lock-in amplifier (521 Hz, 3-20 mV modulation) was used to acquire d$I$/d$V$ spectra. The STM tip was navigated to samples using the capacitance-guiding technique[57]. The STM images were processed with WSxM software.

**Theoretical calculation**

Our first-principles calculations are based on density functional theory (DFT) as implemented in the open-source package for material explorer (OpenMX)[58] and Vienna *ab* initio Simulation Package (VASP)[59]. For the tMoTe$_2$ with twist angle 2.65° calculated by OpenMX, we adopt the relaxed structure with moiré wavelength 76.23 Å and get the electronic structure under self-consistent criterion of $7 \times 10^{-4}$ eV. In the DOS and LDOS calculation in tMoTe$_2$, an $81 \times 81 \times 1$ Monkhorst-Pack $k$-point mesh is applied to sample the Brillouin zone (BZ) and the gaussian smearing of energy is chosen to be 0.007 eV. The valley-resolved DOS is generated by considering the summation of projected components of in-plane atomic orbitals as K valley's contribution and summation of out-of-plane ones as Γ valley's contribution. For the bilayer AB-stacking MoTe$_2$ calculated by VASP, we adopt an in-plane lattice constant of 3.52 Å and apply the van der Waals (vdW) corrections of DFT-dDsC. The ionic potential is treated using the projector augmented wave (PAW) method. Exchange-correlation interactions are handled with the Perdew-Burke-Ernzerhof (PBE) functional within the generalized gradient approximation (GGA). We employ a plane-wave basis set truncated at a cutoff energy of 600 eV. Convergence thresholds are meticulously set at 0.001 eV/Å for forces and $10^{-6}$ eV for total energy. To sample the three-dimensional BZ, we use an $11 \times 11 \times 1$ Monkhorst-Pack k-point mesh.

To mimic the effects of electric field, we employ a continuum model that incorporates first-harmonic, second-harmonic, and displacement field terms to capture the splitting of MX and XM states. The K and K' valley are connected by time-reversal symmetry, making it sufficient to analyze with one valley. Here, we derive the two-band $\boldsymbol{k} \cdot \boldsymbol{p}$ Hamiltonian as:

$$\hat{H} = \begin{bmatrix} -\dfrac{(k - K_t + eA)^2}{2m^*} + \Delta_t(r) - \dfrac{\epsilon}{2} & \Delta_T(r) \\ \Delta_T^\dagger(r) & -\dfrac{(k - K_b - eA)^2}{2m^*} + \Delta_b(r) + \dfrac{\epsilon}{2} \end{bmatrix}$$

where $\Delta_{t/b}(r)/\Delta_T(r)$ denotes the intralayer/interlayer moiré potential term, and $A$ is the strain-induced gauge field. To mimic the effect of electric field, we introduce the displacement field term $\epsilon$. The intralayer and interlayer moiré potential terms can be obtained after the Fourier transformation:

$$\Delta_t(r) = 2V_1 \sum_{i=1,3,5} \cos(g_i^1 \cdot r + l\phi_1) + 2V_2 \sum_{i=1,3,5} \cos(g_i^2 \cdot r)$$

$$\Delta_T = w_1 \sum_{i=1,2,3} e^{-iq_i^1 \cdot r} + w_2 \sum_{i=1,2,3} e^{-iq_i^2 \cdot r}$$

$$A(r) = A(a_2 \sin(G_1 \cdot r) - a_1 \sin(G_3 \cdot r) - a_3 \sin(G_5 \cdot r))$$

Here, $G_{1,3,5}$ and $a_{1,2,3}$ are the moiré reciprocal vectors in reciprocal space and real space. $k$ is the momentum measured from the $\Gamma$ point of a single-layer MoTe$_2$, $K_t/K_b$ represents the momentum of the top/bottom layer. The terms $g_i^1/g_i^2$ and $q_i^1/q_i^2$ represent the nearest/second-nearest distances of plane wave bases in the same and different layers.

By fitting the density functional theory band structures, we get the following parameter [28]: $m^* = 0.62\ m_e$, $V_1 = 10.3$ meV, $V_2 = 2.9$ meV, $w_1 = -7.8$ meV, $w_2 = 6.9$ meV, $\varphi_1 = -75°$, $\Phi/\Phi_0 = 0.737$. Here, $m^*$ is the electron's effective mass. $V_1/V_2$ and $w_1/w_2$ are Fourier components of the intralayer and interlayer strength. $\Phi/\Phi_0$ represents dimensionless flux, which quantifies the flux in a moiré unit cell in units of the quantum flux. Employing these parameters, we diagonalize the Hamiltonian and calculate the density of states (DOS) according to:

$$D(E) = \frac{N_e}{(2\pi)^2} \sum_n \int_{BZ} \delta(E - \epsilon_{n,\mathbf{k}}) d^2 k$$

where $N_e$ represents the band occupancy and $\delta(E - \epsilon_{n,\mathbf{k}})$ is the delta function centered at the energy $E$. To approximate the delta function, we use a Gaussian broadening:

$$\delta(E - E_n) \approx \frac{1}{\sqrt{2\pi\sigma^2}} \exp\left(-\frac{(E - E_n)^2}{2\sigma^2}\right)$$

Here, $\sigma$ is the smearing parameter, and we set to 5 meV during the calculation.

**Acknowledgement**

This work is supported by the National Key R&D Program of China (Nos. 2022YFA1405400, 2022YFA1402401, 2022YFA1402404, 2021YFA1401400, 2022YFA1402702, 2021YFA1400100, 2020YFA0309000), the National Natural Science Foundation of China (Nos. 22325203, 12350403, 12174249, 12174250, 12141404, 92265102, 12374045, 92365302, 92265105, 92065201, 12074247, 12174252, 22272050, 21925201), the Innovation Program for Quantum Science and Technology (Nos. 2021ZD0302600 and 2021ZD0302500), the Natural Science Foundation of Shanghai (No. 22ZR1430900). W.S., T.L. and X.L. acknowledge the Shanghai Jiao Tong University 2030 Initiative Program. X.L. acknowledges the Pujiang Talent Program 22PJ1406700. T.L. acknowledges the Yangyang Development Fund. Y.Z. acknowledges support from AI-Tennessee and Max Planck partner lab grant on quantum materials. C.L. acknowledges China Postdoctoral Science Foundation (No. GZB20230422). N.M. acknowledges the support from the Alexander von Humboldt Foundation. K.W. and T.T. acknowledge support from the JSPS KAKENHI (Nos. 21H05233 and 23H02052) and World Premier International Research Center Initiative (WPI), MEXT, Japan.

**Author contributions**

S.W. and T.L. designed and supervised the experiments. Y.G. and S.J. fabricated the devices. Y.L. and C.L. performed the STM/STS measurements. L.L., D.G., Y.L., H.Z., C.L., J.J., S.W., T.L. and Y.Z. analyzed the data. T.B., N.M., W.D. and Y.Z. performed theoretical studies. K.W. and T.T. grew the bulk hBN crystals. Y.L., S.W., T.L. and Y.Z. wrote the manuscript. All authors discussed the results and commented on the manuscript.

**Competing interests**

The authors declare no competing financial interests.

**Data and Code Availability**

All data and code are available from the corresponding authors upon reasonable request.

**Materials & Correspondence**

Correspondence and requests for materials should be addressed to Can Li, Yang Zhang, Tingxin Li, and Shiyong Wang.

**Extended Data Figures**

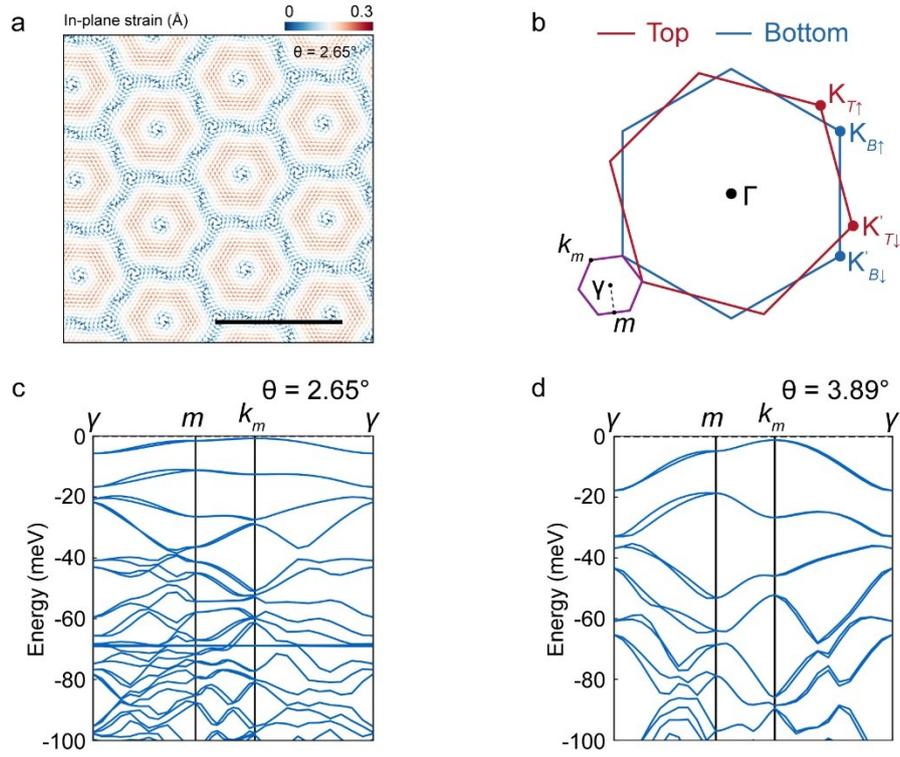

**Extended Data Fig. 1. DFT calculations of tMoTe₂ a,** Calculated lateral strain distribution in 2.65° tMoTe₂. Scale bar represents 10 nm. **b**, Schematic of Brillouin zones: The Brillouin zones of the twisted bilayer MoTe₂ (red/blue) and the mini-Brillouin zone of the emergent moiré superlattice (purple). **c,d**, Calculated band structures, incorporating lattice relaxations, for twist angles of 2.65° (c) and 3.89° (d).

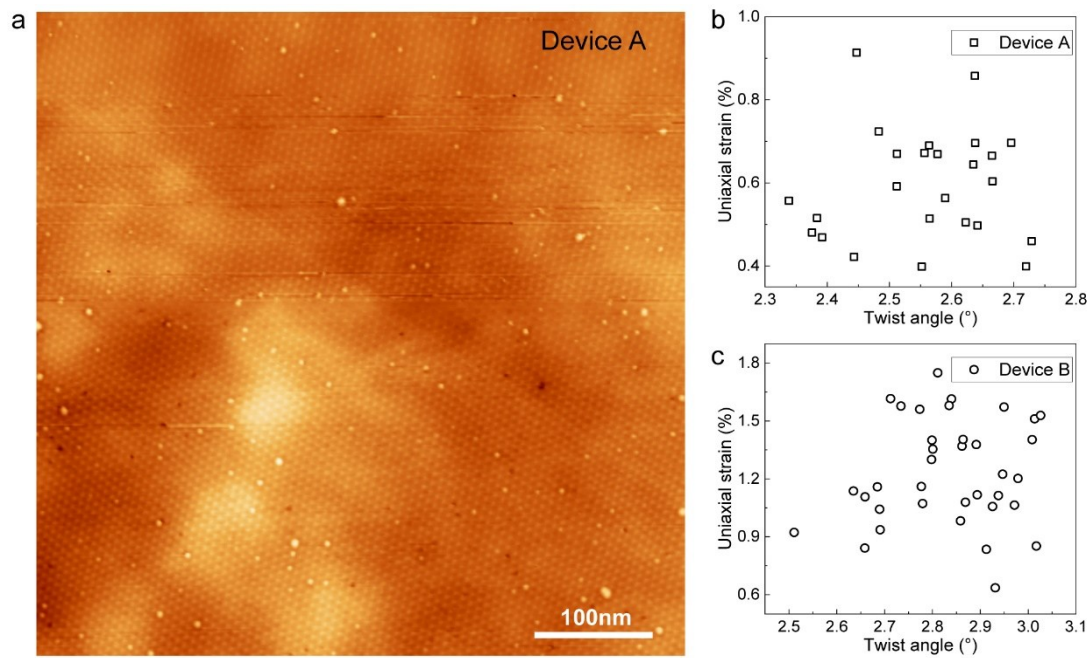

**Extended Data Fig. 2. Large scale STM image of Device A. a**, STM topography of tMoTe$_2$ (**V**$_{bias}$ = -2.0 V and **I** = 145 pA). The sample is micrometer-size clean with few point defects. **b,c,** The analyzed uniaxial strain as a function of twist angle of device A **(b)** and device B **(c)**.

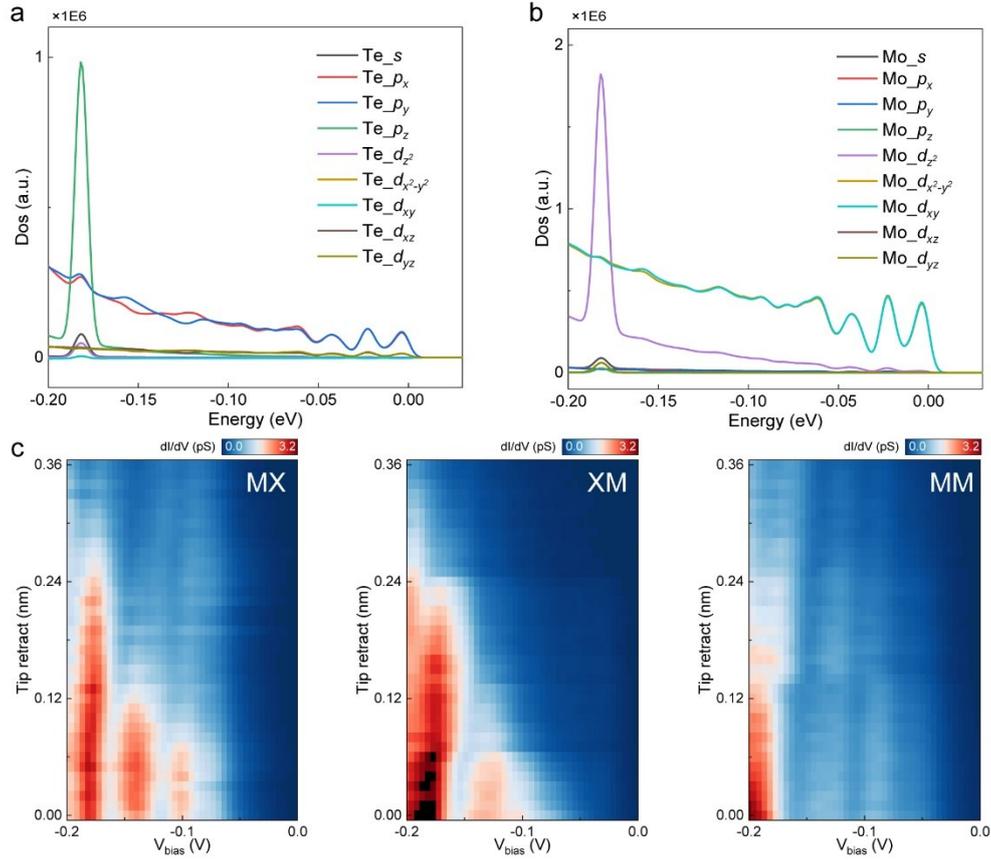

**Extended Data Fig. 3. Orbital characters of Γ and K bands in tMoTe$_2$. a,b**, DFT calculated orbital contributions of Γ and K bands of tMoTe$_2$ with a twist angle of 3.89°. Γ-valley bands originate from the $d_{z^2}$ orbitals of Mo and Te, which possess out-of-plane orbital characters and therefore decay slowly outside the MoTe$_2$ layer. The K-valley states originate from the in-plane $d_{xy}$ orbitals of Mo and the $p_x$ and $p_y$ orbitals of Te. These wavefunctions have large in-plane orbital characters and decay rapidly outside the MoTe$_2$ layer. **c,** STS d$I$/d$V$ spectra plots taken at MX, XM, and MM sites with varied tip-sample separations. Bias modulation: 5 mV. (**θ ≈ 3.6°, Setup 2, Device C**).

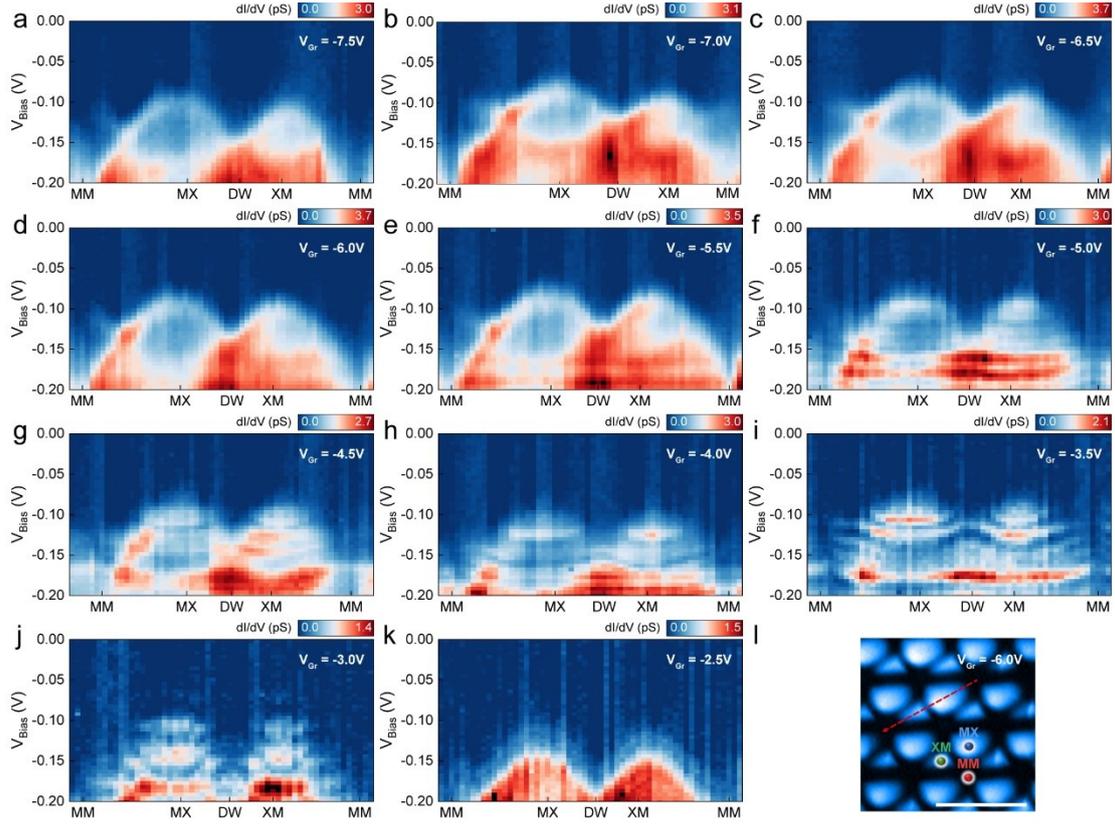

**Extended Data Fig. 4. Modulating moiré flat bands from K valleys by displacement electric fields. a-k**, STS d$I$/d$V$ spectra measured along the dashed line in **(l)** at different graphite gate voltages (θ ≈ **3.3°, Setup 2, Device E**). These spectra demonstrate that the K-valley flat bands are highly tunable via the applied displacement fields. **l**, Constant-height current image revealing the emergence of two distinct triangular lattices at the MX and XM sites, respectively. Scale bars: 10 nm. In the graphite gate range of -2.5 V to -4.0 V, the MX/MX bands are degenerate, and isolated peaks corresponding to different moiré flat bands can be visualized.

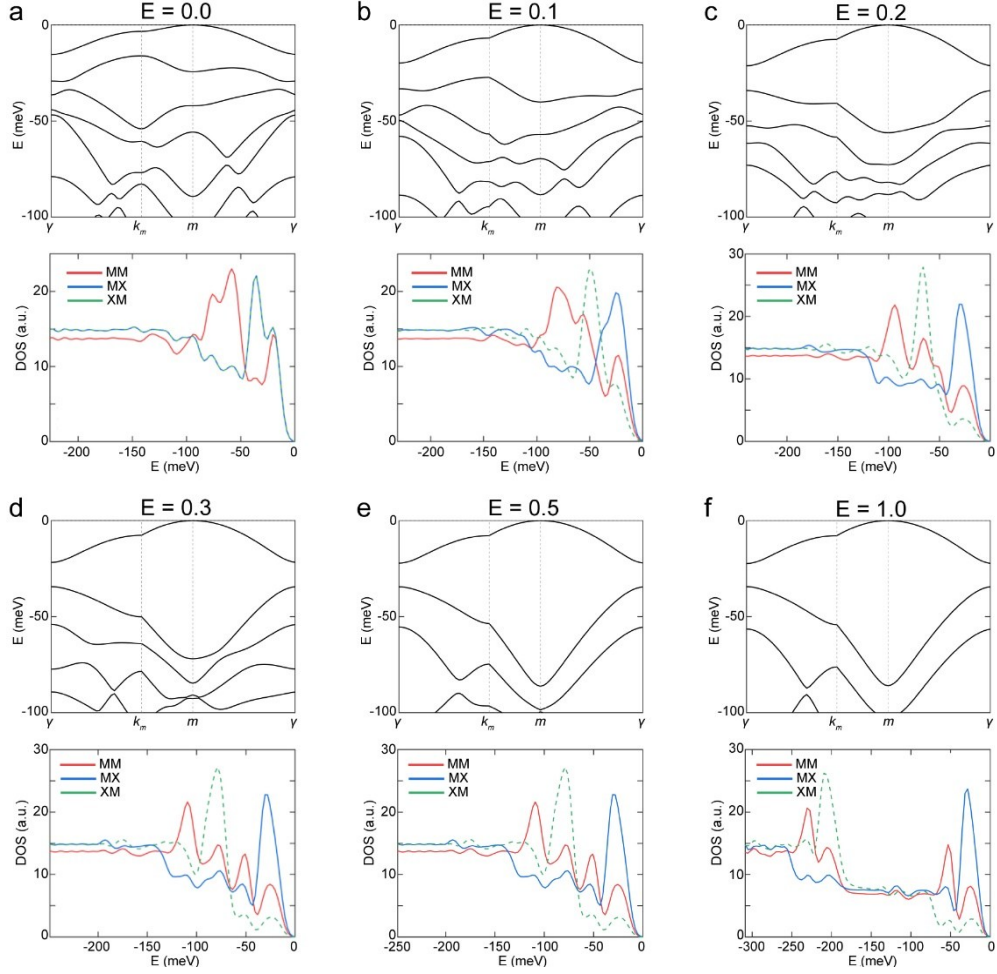

**Extended Data Fig. 5. Continuum Model Calculations of Band Structure and DOS Spectra of 3.65º tMoTe$_2$.** The band structure is highly sensitive to displacement electric field, which gradually lifts the degeneracy of the XM/MX states. At zero field, the DOS for the XM and MX states are identical and lie closer to the Fermi level than those of the MM sites. As the displacement field increases, this degeneracy is broken, leading to an energy splitting between the corresponding DOS peaks that grows with the field strength.

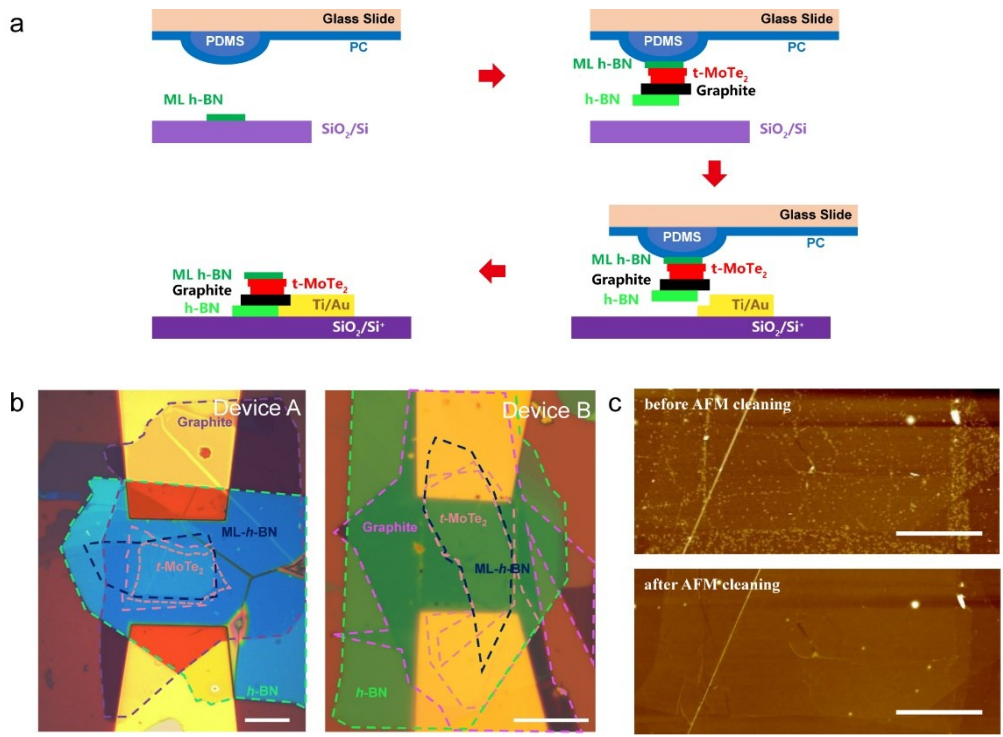

**Extended Data Fig. 6. Device fabrication for Setup 1. a,** Schematic of the device fabrication processes. **b,** Optical micrographs of the tMoTe$_2$ device A and device B (from left to right). Dashed lines highlight: pink/blue for tMoTe$_2$/hBN monolayers; purple/green for bottom graphite/hBN flakes. **c,** AFM topography of the tMoTe$_2$ device B before and after contact-mode AFM cleaning. Scale bars: 10 μm (**b**); 5 μm (**c**).

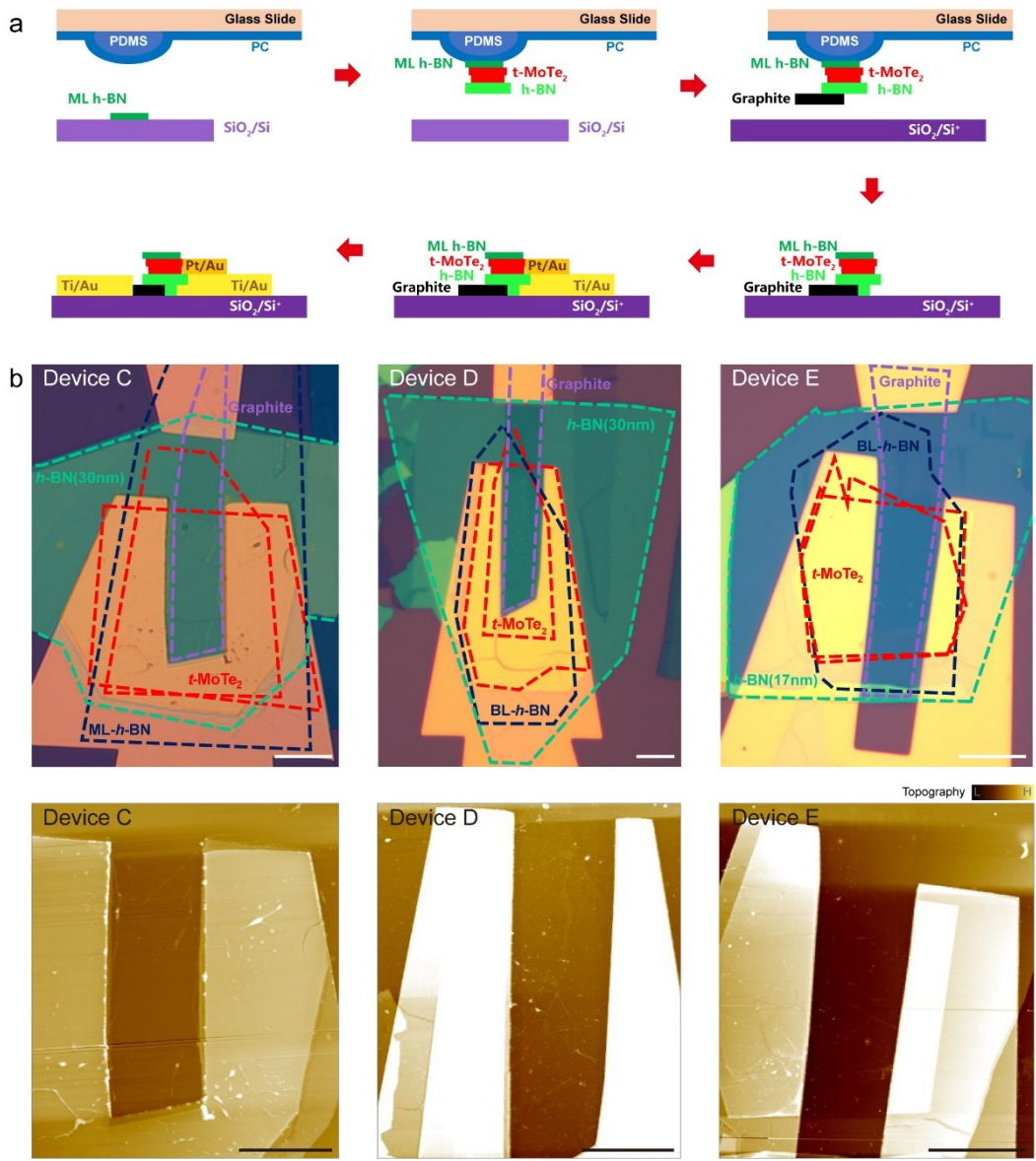

**Extended Data Fig. 7. Device fabrication for Setup 2. a,** Schematic of the device fabrication processes. **b,c,** Optical micrographs and AFM images of the tMoTe₂ device C, D and E (from left to right). Colored lines highlight: red/blue for tMoTe$_2$/hBN monolayer/bilayer; purple/green for bottom graphite/hBN flakes. Scale bars are 10 μm.

28